\documentclass[12pt]{iopart}
\usepackage{iopams}
\usepackage{graphicx}

\begin{document}

\newcommand{\be}{\begin{equation}}
\newcommand{\ee}{\end{equation}}

\title{Polarized thermal emission by thin metal wires}

\author{ Giuseppe Bimonte\dag, Luca Cappellin${}^*$, Giovanni Carugno\ddag, Giuseppe Ruoso${}^+$, Daniela Saadeh${}^*$}
%\thanks{}
\address{\dag
Dipartimento di Scienze Fisiche Universit\`{a} di Napoli Federico
II Complesso Universitario MSA Via Cintia
I-80126 Napoli Italy, and INFN, Sezione di Napoli, Napoli, ITALY\\
}

\address{\ddag INFN, Sezione di Padova, Via Marzolo 8, 35131 Padova,
Italy}
\address{${}^+$ INFN, Laboratori Nazionali di Legnaro, Viale dell'Universit\`a 2, 35020 Legnaro (Padova),
Italy}
\address{${}^*$ Scuola Galileiana di Studi Superiori, Via Giovanni Prati 19, 35122 Padova, ITALY}

\ead{bimonte@na.infn.it}

\begin{abstract}
We report new   measurements of the linear polarization of thermal
radiation emitted by incandescent thin tungsten wires, with
thicknesses ranging from five to hundred microns. Our data show
very good agreement with theoretical predictions, based on
Drude-type fits to measured optical properties of tungsten.

\end{abstract}

\maketitle

\section{Introduction}

Thermal radiation by incandescent bodies has been the subject of
intense theoretical and experimental investigations for over a
century now. While the laws of thermal emission by a blackbody are
at the very roots of Quantum Theory \cite{planck}, the richness of
phenomena involved in thermal emission has been fully realized
only quite recently. Perhaps, the most  interesting progress has
been the realization that thermal radiation may exhibit a
significant degree of spatial and temporal coherence, in seeming
contradiction with one's idea that thermal emission is an
incoherent phenomenon. Among the most recent findings, we mention
as an example the remarkable degree of spatial coherence  of the
radiation of a hot body in the near-field region \cite{carminati}.
Coherence features of thermal radiation are also at the basis of
recent attempts to modify or tailor the profile of thermal
emission by metallodielectric surfaces, with subwavelength
patterns, that are of great importance in applied physics and
engineering (see Refs.\cite{laroche,chan,klein,dahan,biener,au}
and references therein).

Another striking example of this sort was  discovered quite some
time ago by $\ddot{\rm O}$hman \cite{ohman}, who observed that
visible radiation from a hot thin metallic wire shows a
significant polarization. Using incandescent tungsten filaments
with a thickness of a few microns, this author found  a
polarization in the direction orthogonal to the wire as high as 28
per cent, in the red region of the spectrum. In a related attempt
to explain these preliminary findings, Agdur et al. \cite{agdur}
investigated in detail the scattering and absorption of light by
thin metal wires, using silver wires having a diameter down to
2000 \AA. The data were compared with a simple theoretical model,
where the "plasma" properties of the metal were taken into
account, showing good agreement with the measurements. Even though
scattering and absorption data are related theoretically, via
Kirchhoff's law, to thermal emission, the authors of
Ref.\cite{agdur}  could not perform accurate measurements of the
degree of polarization of the radiation emitted by the wires, due
to technical difficulties, and only report that a polarization of
about fifty per cent was found in the case of silver filaments
with a diameter of about 0.8 microns.

After these early findings, several authors have recently
investigated the polarization features of the thermal radiation
emitted by a number of sources with different designs, like
platinum microwires \cite{klein,au}, semiconductor layers placed
in an external magnetic field \cite{kollyukh}, SiC lamellar
gratings \cite{marquier}. In this paper, we report new
measurements of the "linear polarization" (see Sec. 2) of thermal
radiation emitted by individual incandescent tungsten wires, with
thicknesses between five and hundred microns. Our work is closely
related to Refs.\cite {klein,au}, which report measurements of the
polarization and angular distribution of thermal radiation from
individual antenna-like, thin film platinum microwires, heated at
a temperature of 900 K. While the quantity that we measure to
characterize the polarization of the thermal radiation is
essentially the same as the "extinction ratio $E$" measured in
\cite{klein,au}, two differences between our work and
Refs.\cite{klein,au} should be stressed. Apart from the fact that
we use tungsten instead of platinum, which allows us to work in
the visible region of the spectrum, the main difference is in the
relative magnitude of the wires thickness, as compared to the
wavelength of the observed radiation. While the lateral extent of
the microwires of Refs. \cite{klein,au} is in fact smaller than
(or comparable to) the wavelengths of the infrared radiation
observed there, we are quite in the opposite situation, since our
wires are always much thicker than the wavelengths that we
observe. As it will be seen in greater detail in the next Section,
the polarization features of the thermal radiation are quite
opposite in the two regimes: while for very thin wires, as
reported in \cite{klein,au}, the thermal radiation is polarized in
a direction {\it parallel} to the wire, in the case of thicker
wires the situation is reversed, and the radiation is now
polarized in the direction {\it orthogonal} to the wire, the
crossover occurring for wavelengths roughly equal to the
circumference of the wire. This latter case is the one originally
reported in Ref. \cite{ohman}, and it is the one explored in the
present paper.

The paper is organized as follows: in section 2 we derive the
theoretical expression for the linear polarization of thermal
emission by a wire. In section 3 we describe our apparatus and
present our measurements, while in section 4 the experimental data
are compared with theoretical predictions. In Section 5 we draw
our conclusions, and finally in the Appendix we present the
detailed expression of the Drude-type fit to the optical data of
tungsten, as given in Ref.\cite{roberts}, that we used for our
numerical computations.

\section{Theory}

\subsection{General equations}

We  model the wire as a homogeneous circular cylinder ${\cal C}$
of length $l$ and radius $a$. The observation point  is placed at
a distance $r$ from the wire, in the plane passing trough the mid
point of the wire and orthogonal to it. It is further assumed that
$a$, $l$ and $r$ satisfy the conditions \be a^2 \ll \lambda \,r
\ll l^2\;, \ee where $\lambda$ is the wavelength of radiation.
Under this assumption, the radiation field at the observation
point coincides with the far-field for an infinitely long
cylinder. The material constituting the wire is described as a
homogenous dielectric with a complex permittivity
$\epsilon(\omega)$ depending on the frequency $\omega$, and
obviously on the temperature $T$ of the wire. As we consider
non-magnetic materials, we shall set to one the magnetic
permeability $\mu$.

Cylindrical symmetry of the system permits to introduce TE and TM
modes of the electromagnetic field \cite{jacks}: TE modes have
their electric field ${\bf E}$ in the plane orthogonal to the
cylinder axis, which we take to coincide with the $z$ direction.
On the contrary, TM modes have their magnetic field ${\bf B}$
orthogonal to the cylinder axis. At large distances from the wire,
the electric field is orthogonal to the line of sight, and for TE
modes it is orthogonal to the wire, while for TM modes it is
parallel to it. The $z$-components of the electric and magnetic
fields, $E_z$ and $B_z$, can be taken as independent fields, from
which all other components of ${\bf E}$ and ${\bf B}$ can be
obtained using Maxwell equations.   TE modes are then
characterized by the condition $E_z^{({\rm TE})}=0$, while for TM
modes we have $B_z^{({\rm TM})}=0$. Therefore TE and TM modes are
labelled by $B_z$ and $E_z$, respectively. Outside ${\cal C}$,
both $B_z$ and $E_z$ satisfy the Helmholtz equation: \be
-(\nabla^2+k^2)E_z= -(\nabla^2+k^2)B_z=0\,,\ee where $k=\omega/c$.
We collect $B_z$ and $E_z$ in a two-dimensional vector ${\bf u}$:
\be {\bf u}\equiv \left(\begin{array}{c}
 B_z \\  E_z \\
\end{array}\right)\,.\ee
Since we   only observe radiation that is emitted in a narrow
solid angle around a direction orthogonal to the $z$ axis, we can
just consider fields that propagate in the $(x,y)$ plane and are
thus  independent of the $z$-coordinate. For normal incidence, TE
and TM modes do not mix under scattering by the wire and
therefore, outside ${\cal C}$, ${\bf u}$ is a superposition of
partial waves ${u}_{m}^{(\alpha)}$ with definite polarization
$\alpha$ and angular momentum $m$ along the $z$-axis: \be
{u}_{m}^{(\alpha)}=\left(\,H_m^{(2)}(k \,r)+ { {\cal
S}}_m^{(\alpha)}(k) \, H^{(1)}(k\, r) \right)e^{i m \phi}\,, \ee
where $\phi$ is the azimuthal angle, and $H^{(i)}$ are Hankel
functions. The scattering amplitudes ${{\cal S}}_m^{(\alpha)}(k)$
measure the response of the wire to a unit amplitude incoming wave
with wave-vector $k$, polarization ${\alpha}$ and angular momentum
$m$ along the wire. It should be noted that, with the wire absent,
${{\cal S}}_m^{(\alpha)}(k)\rightarrow 1$, and then the partial
waves ${u}_{m}^{(\alpha)}$ would reduce to $2 \,J_m(k\,r) \exp(im
\phi)$, the partial wave solution regular along the $z$-axis.

By use of Kirchhoff's law, one easily finds that the emissivity
$e^{(\alpha)}(k)$ of the wire for polarization $\alpha$ is: \be
e^{(\alpha)}(k)= \sum_{m=-\infty}^{\infty}(1-\vert {{\cal
S}}_m^{(\alpha)}(k)\vert^2)\;.\ee
%where  $A$ is a geometric factor
%independent of the polarization $\alpha$.
It is convenient to reexpress the above formula for the emissivity
in terms of the so-called transition amplitudes ${\cal
T}_m^{(\alpha)}(k)$ defined as: \be {\cal T}_m^{(\alpha)}(k) =
\frac{1}{2}(1-{\cal S}_m^{(\alpha)}(k))\;.\ee We then obtain: \be
e^{(\alpha)}(k)=4 \sum_{m=-\infty}^{\infty}[\,{\rm Re}({\cal
T}_m^{(\alpha)}(k))-\vert {{\cal
T}}_m^{(\alpha)}(k)\vert^2)\,]\,.\ee The explicit expressions for
the transition amplitudes ${\cal T}_m^{(\alpha)}(k)$ can be easily
obtained by solving the scattering problem for a homogeneous
dielectric cylinder \cite{jacks}. The result is: \be {\cal
T}_m^{({\rm TE})}(k)=\frac{J'_m(n k a)\,J_m(k a)- n\,J'_m(k
a)\,J_m(n k a)} {J'_m(n k a)\,H_m^{(1)}(k a)- n\,J_m(n k
a)\,H_m^{(1)\,'}(k a)}\,,\ee \be {\cal T}_m^{({\rm
TM})}(k)=\frac{J_m(n k a)\,J_m'(k a)- n \,J'_m(n k a)\,J_m(k a)}
{J_m(n k a)\,H_m^{(1)\,'}(k a)- n\,J'_m(n k a)\,H_m^{(1)}(k
a)}\,.\ee In the above formulae, a prime denotes differentiation,
while $n=\sqrt{\epsilon}$ is the (complex) refraction index.
Following the terminology of Ref.\cite{smith}, we define the {\it
linear polarization} $P(k)$   of the thermal emission of the wire
by the formula: \be P(k)=\frac{e^{({\rm TE})}(k)-e^{({\rm
TM})}(k)}{e^{({\rm TE})}(k)+e^{({\rm TM})}(k)}\,\label{pol}\ee
with the positive sign corresponding to polarization in the
direction orthogonal to the wire.   The quantity $P(k)$ above is
the observable considered in Refs.\cite{ohman,agdur} and it is
basically the same as the extinction ratio $E$ considered in
Refs.\cite{klein,au} ($P=-E/(2+E)$). We should remark  that the
quantity $P$ does not provide a complete characterization of the
degree of polarization of the radiation, which is properly
described in terms of the Stokes parameters \cite{born}. A
complete determination of the Stokes parameters requires
observation of the circular polarization of the emitted radiation,
but unfortunately our apparatus does not permit it. We therefore
content ourselves with the partial information contained in the
linear polarization $P$ above. In Figure 1 we plot the theoretical
prediction for $P$ as given by Eq. (\ref{pol}), for a tungsten
wire, as a function of $\log (2 \pi a/\lambda)$, for $\lambda=.5\;
\mu$m. The curve was computed using the Drude-like fit to the
optical data for tungsten quoted in Ref.\cite{roberts}, for a wire
temperature $T=2400$ K (see Appendix for details). The most
striking feature seen in Figure 1 is the change of sign of $P$
that occurs as the thickness of the wire is decreased: while for
thick wires the radiation is polarized in the direction orthogonal
to the wire, for wires having a thickness (or better to say a
circumference $2\, \pi \,a$) comparable to or smaller than the
wavelength $\lambda$, the axis of polarization aligns with the
wire. We remark that the former behavior is the one reported for
the first time in \cite{ohman,agdur}, while the latter one was
observed recently in \cite{klein,au} using very thin platinum
microwires. Another remarkable feature of $P$ as a function of $a$
is not clearly visible from Figure 1, and can be better seen from
Figure 4 below: in the thick wire regime, $P$ is not a
monotonically increasing function of $a$, and it displays a
maximum for a definite value of $a$. We have not been able to find
a simple explanation of this feature of the curve.

The quantity that we actually measure  really is an average
$\mathbb{P}$ of  $P(k)$ over the wavelengths that get through a
polarizing filter placed between the wire and the detector. The
transmission efficiency of the filter is characterized by a
transmission coefficient $\chi(\lambda)$, comprised between zero
(no transmission) and one (full transmission). The average linear
polarization $\mathbb{P}$ can then be written as: \be
\mathbb{P}=\frac{{\bar e}^{({\rm TE})}-{\bar e}^{({\rm
TM})}}{{\bar e}^{({\rm TE})}+{\bar e}^{({\rm
TM})}}\,,\label{avpol}\ee where \be {\bar e}^{(\alpha)}=
\frac{1}{N}\int_{0}^{\infty} d \lambda\, \chi(\lambda)\,
E(\lambda,T)\,e^{(\alpha)}(2 \pi/\lambda) \;.\ee In this equation,
$E(\lambda)$ is Planck formula (expressed in terms of the
wavelength $\lambda$): \be E(\lambda,T)=\frac{2 \pi h
\,c^2}{\lambda^5} \frac{1}{e^{\lambda_T/\lambda}-1}\;,\ee where
$\lambda_T=h c/(k_B T)$, and $N$ is a normalization constant
$N=\int_0^{\infty} d \lambda \chi(\lambda) E(\lambda,T)$.
\begin{figure}
\includegraphics{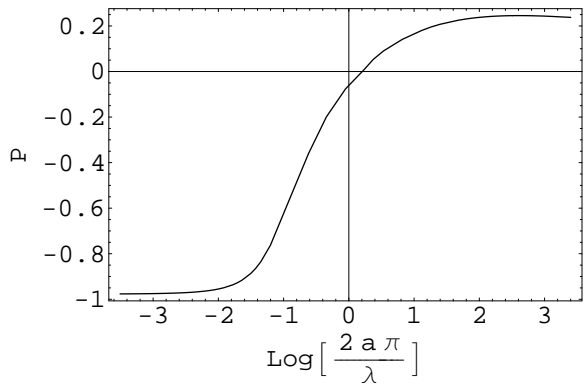}% Here is how to import EPS art
\caption{\label{plots}  Plot of the linear polarization $P$ (see
text for explanation) for a tungsten wire of radius $a$,  versus
$\log (2 \pi a/\lambda)$. The curve is computed for a fixed
wavelength $\lambda=.5 \,\mu$m, using the Drude-like fit to the
optical data of tungsten for $T=2400$ K (see Appendix).}
\end{figure}

\subsection{The limit of thick wires}

In the limit of thick wires, $a \gg \lambda$, it is possible to
derive a very simple expression for the linear polarization $P$ of
the radiation emitted by the wire in a direction orthogonal to $z$
\cite{bertilone}. When $a \gg \lambda$, diffraction effects become
negligible, and we can regard the surface of the wire as locally
flat.  Let us say, for definiteness, that the line of sight
coincides with the $x$ direction.  Then, the emissivity of the
wire in the direction $x$, for polarization $\alpha$, is easily
found to be: \be e^{(\alpha)}(k)=\frac{1}{2} \int_{-\pi/2}^{\pi/2}
d \phi \cos(\phi)\; e^{(\alpha)}_{\rm plane}(k,
\phi)\,,\label{aver}\ee where $e^{(\alpha)}_{\rm plane}(k, \phi)$
is the emissivity for a flat surface made of the same material, in
the direction forming an angle $\phi$ with the normal to the
plane. On the other hand, from Kirchhoff's law we have: \be
e^{(\alpha)}_{\rm plane}(k, \phi)=1-\vert
R^{(\alpha)}(k,\phi)\vert ^2\,,\label{kpl}\ee where
$R^{(\alpha)}(k,\phi)$ are the familiar Fresnel reflection
coefficients for radiation incident at an angle $\phi$. Using Eqs.
(\ref{aver}) and (\ref{kpl}) in the expression  of $P$, Eq.
(\ref{pol}), we obtain: \be P \simeq \frac{\int_{-\pi/2}^{\pi/2} d
\phi \,\cos \phi\,(\vert R^{({\rm TE})}(k,\phi)\vert ^2-\vert
R^{({\rm TM})}(k,\phi)\vert ^2)}{\int_{-\pi/2}^{\pi/2} d \phi
\,\cos \phi \,(2-\vert R^{({\rm TE})}(k,\phi)\vert ^2-\vert
R^{({\rm TM})}(k,\phi)\vert ^2)}\;.\ee With our choices of
polarization, the explicit expressions for $R^{(\alpha)}(k,\phi)$
are: \be R^{({\rm TE})}(k,\phi)=\frac{\epsilon \cos
\phi-\sqrt{\epsilon-1 +\cos^2 \phi}}{\epsilon \cos
\phi+\sqrt{\epsilon-1 +\cos^2 \phi}}\;,\ee \be R^{({\rm
TM})}(k,\phi)=\frac{\cos \phi-\sqrt{\epsilon-1 +\cos^2 \phi}}{\cos
\phi+\sqrt{\epsilon-1 +\cos^2 \phi}}\;.\ee

\section{Experiments}

\begin{figure}[htb]
\includegraphics{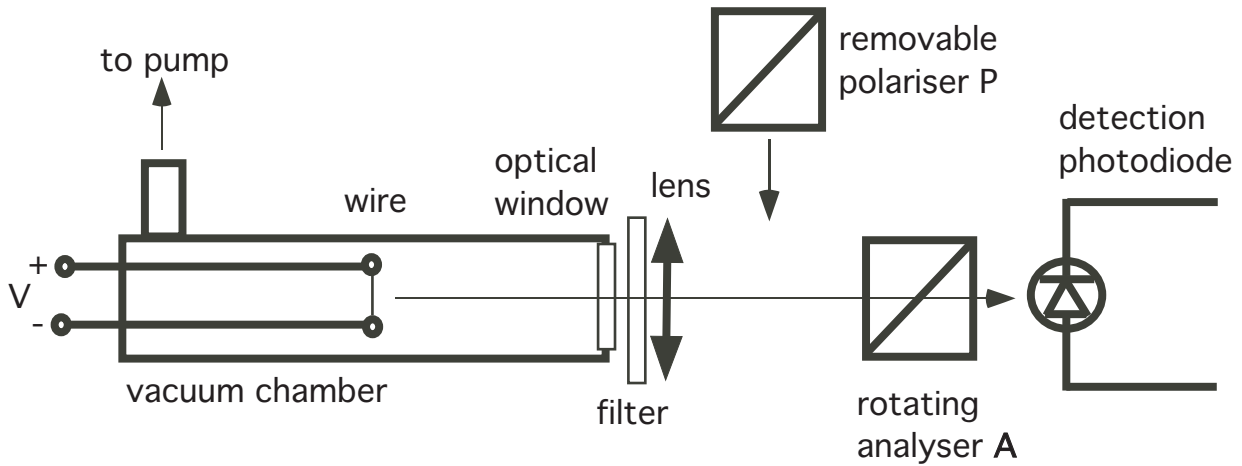}% Here is how to import EPS art
\caption{\label{schema} Drawing (not to scale) of the experimental apparatus.}
\end{figure}

The experimental apparatus (see figure \ref{schema}) consists of a vacuum cylindrical chamber in which the metal
wire is positioned perpendicular to the chamber axis. The chamber is 52 cm long and has a 2.5 cm diameter, and a
pumping system provides a working residual pressure about $10^{-6}$ mbar. The 7 mm long wire is sustained by two
electrical feedthroughs providing the flowing current to set the wire incandescent. An optical window allow for
light emitted by the wire to exit from the vacuum region. The tube internal surface is mat and painted with aquadag
in order to prevent reflections: this way, light emerges from the tube roughly parallel to its axis. This light,
after passing a bandpass filter is imagined onto a detection photodiode by a lens. In order to study the polarization
properties of emitted light, a pair of polaroid polarizers have been used. These polarizers (Edmund Optics TECHSPEC)
work only in a limited bandwidth centered in the visible domain. Since light emitted from the wire lies mainly in
the infrared spectral region, the bandpass filter (Thorlabs model FES750) selects light in the spectral band from 450
to 750 nm. In order to avoid systematics due to residual infrared light impinging on the photodiode, the following
procedure has been used.

Step 1. The rotating analyser {\bf A} is placed on the optical path. Measurements are taken rotating the analyser
and recording the photodiode signal every 0.5 degree. The light intensity on the photodiode can be written as:

\begin{equation}
I_1 =  A_1 \cos^2( \theta - \theta^*)+F_1\;,
\end{equation}

\noindent where $\theta$ is the angular position of the analyser,
$A_1$ measures the maximum amount of polarized light impinging on
the photodiode and $F_1$ is the sum of unpolarized light and of
the residual infrared light. Due to this residual, this single
measurement is not sufficient to determine the linear
polarization, which is given by:

\begin{equation}
\mathbb{P}=\frac{I_P}{I_P+I_U}\;,
\end{equation}

\noindent where $I_P$ and $I_U$ refer to the intensities of
linearly  polarized and unpolarized light emitted by the filament.
The purpose of this step 1 is to determine the angle $\theta^*$ of
maximum transmission for the emitted light (Electric field), which
corresponds to the maximum for $I_1$.

Step 2. Insert the polarizer {\bf P} in the optical path. Two sets of measurements are taken with the analyser,
for the two directions of the polariser {\bf P} as $\theta^*$ and $\theta^*+\frac{\pi}{2}$.

\begin{eqnarray}
I_2^a & = & A_a \cos^2 (\theta-\theta_a) + F_2\,\,\,\,\,\,\,\, ({\rm {\bf P}\,\, set\,\, to\,\, } \theta^*)\\
I_2^b & = & A_b \cos^2 (\theta-\theta_b) + F_2\,\,\,\,\,\,\,\,
({\rm {\bf P}\,\, set\,\, to\,\, } \theta^*+\frac{\pi}{2})\;.
\end{eqnarray}
Now $F_2$ has only the infrared residual, while $A_a$  is
proportional to $I_P$ plus half the unpolarized light $I_U/2$  and
$A_b$ only to $I_U/2.$ It follows then:

\begin{equation}
\mathbb{P}=\frac{A_a-A_b}{A_a+A_b}\;.
\end{equation}
Collected data are fitted using least square analysis. This allows
a precise determination of the coefficients $A_i$ and a control of
the correct positioning of the polarimeter {\bf P}: in all the
measurements the discrepancy between the angles
$\theta_a,\theta_b$ and $\theta^*,\theta^*+\frac{\pi}{2}$, was
kept below 0.5 degree.

\begin{figure}[htb]
\includegraphics{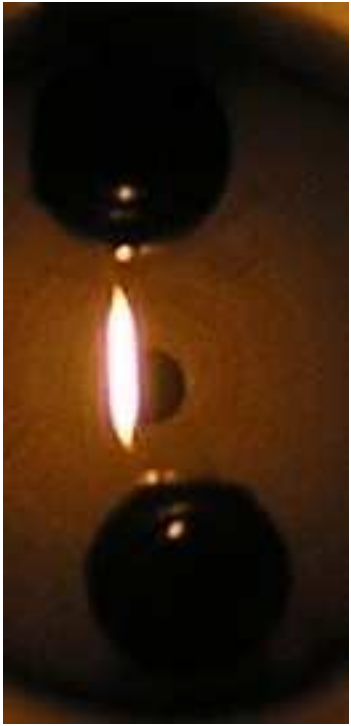}% Here is how to import EPS art
\caption{\label{filo} A photograph of the 5 $\mu$m diameter wire with 10 mA current flowing through it.
The two black circles on the upper and lower ends are the feedthroughs.}
\end{figure}

Measurements have been performed using four pure tungsten wires
provided by LUMA Metall, with diameter 5, 17, 35 and 100 $\mu$m.
For each wire three different values of voltage were applied to
the feedthroughs, to check for possible changes of the degree of
polarization with temperature. To get a rough estimate of the
temperature two different methods have been used: in the first the
resistivity is calculated and compared to tabulated values. In the
second, temperature is deduced by the assumption that  the total
electrical power is converted into radiation. The computed values
are affected by large errors, mainly due to the fact that
temperature is not uniform along the wire, but rather follows a
sort of flat top profile. This can be seen for example from the
picture of Figure \ref{filo}. The estimates for the temperature
show that values are in the range 2600 - 3200 K. For all the wires
a range of temperature of 300 -- 400 K is spanned by varying the
voltage.

\begin{table}[htdp]
\caption{Summary of the measurements.}
\begin{center}
\begin{tabular}{||c|c|c|c|c||}
\hline \hline
diameter ($\mu$m) & voltage (V) & current (A) & $\mathbb{P}$ & avg($\mathbb{P}$)\\
\hline \hline
5 & 2.5 & 8.8 10$^{-3}$ & 0.2416 & 0.241 $\pm$ 0.005 \\
\hline
 & 3.0 & 9.6 10$^{-3}$ & 0.2353 &  \\
\hline
 & 3.9 & 10.9 10$^{-3}$ & 0.2446 & \\
\hline \hline \hline
17 & 1.7 & 9.6 10$^{-2}$ & 0.2179  & 0.221 $\pm$ 0.003\\
\hline
 & 2.0 & 0.106 & 0.2237 & \\
\hline
 & 2.4 & 0.117  & 0.2219 & \\
\hline \hline \hline
35 & 1.1 & 0.309  & 0.2119 &  0.208 $\pm$ 0.003\\
\hline
 & 1.2 & 0.32  & 0.2059  &   \\
\hline
 & 1.4 & 0.343  & 0.2078  & \\
\hline \hline \hline
100 & 2.0 & 1.81  & 0.2028 & 0.199 $\pm$ 0.004 \\
\hline
 & 2.3 & 1.96  & 0.1956  &\\
\hline
 & 2.45 & 1.98  & 0.1981 &  \\
\hline \hline
\end{tabular}
\end{center}
\label{risultati}
\end{table}%

Table \ref{risultati} lists the obtained results. Measurements accuracy affects the determination of each value of
$\mathbb{P}$ in a negligible manner: the largest absolute error resulting 0.001 (relative error less than  0.5 \%),
much smaller than the spread of the values for each single wire. The average value has then been taken as the
arithmetic mean with the error the standard deviation. The polarization direction (Electric field)  has been
found orthogonal to the wire for all measurements.

\section{Comparison with theory}

For the purpose of comparing our data with the theoretical value
of $\mathbb{P}$, the following choices were made. We approximated
the transmission coefficient $\chi(\lambda)$ of the polarizer by a
stepwise constant function, as follows: \be \chi(\lambda)=\left\{
\begin{array}{c}
  0\;\;\;\;\;{\rm for}\;\lambda<\lambda_1 \\
  \;\;\;\;\;\;\;1 \;\;\;\;{\rm for}\;\lambda_1<\lambda<\lambda_2\\
  0\;\;\;\;\;{\rm for}\;\lambda>\lambda_2 \\
\end{array}\right.\ee
and we took $\lambda_1=0.5$ micron and $\lambda_2=0.75$ micron.
This crude model describes sufficiently well the actual
transmission coefficient of the filter we used.  For the complex
dielectric function $\epsilon(\omega)$ we used the Drude-type
analytical  fits to optical data of tungsten quoted in
Ref.\cite{roberts} (see Appendix for details). This reference
provides the permittivity of tungsten for several temperatures in
the range from 298 K to 2400 K. Unfortunately, no data are
reported for temperatures higher than 2400 K, as it is the case in
our measurements. We remark that the formulae reported in
Ref.\cite{roberts} contain no adjustable free parameters. In
Figure 4, we show our experimental data for three different
thicknesses (full diamonds with error bars) together with the
plots of three theoretical curves for $\mathbb{P}$ as a function
of the wire diameter (in microns). The theoretical curves have
been computed for three different temperatures, $T=298$ K,
$T=1600$ K and $T=2400$ K. We can see clearly that the theoretical
curve computed using  room-temperature optical data is definitely
not consistent with our measurements, while already the curve for
2400 K fits the data rather well. Since the three theoretical
curves show that the polarization decreases by increasing
temperature, it is likely that an even better agreement with the
data would have resulted, had we had at our disposal data for the
permittivity relative to temperatures around 2600-3200 K, or so,
which we expect to be the range of temperatures reached by our
wires.
\begin{figure}
\includegraphics{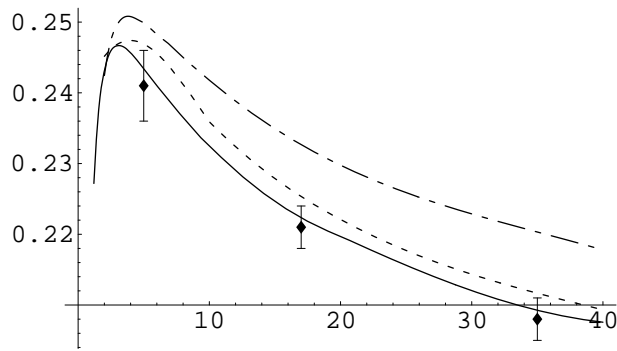}% Here is how to import EPS art
\caption{\label{plots} Experimental data (diamonds with error
bars) and theoretical curves for the average linear polarization.
The wire diameter along the $x$-axis is in microns. The three
theoretical curves displayed have been computed using the
Drude-like fits (with no free parameters) reported in
\cite{roberts} to optical data  of tungsten for three different
temperatures: the solid line is for $T=2400$ K, the dotted line is
for $T=1600$ K, the dot-dashed line is for $T=298$ K.}
\end{figure}
In Table \ref{comp} we quote the detailed experimental data, and the
theoretical prediction for $T=2400$ K.
\begin{table}\begin{tabular}{|c|c|c|c|c|}
  \hline
  diameter ($\mu$m) & 5 & 17 & 35 & 100 \\
  \hline
  $\mathbb{P}_{\rm exp}$ & 0.241$\pm$0.005 & 0.221$\pm$0.003 & 0.208$\pm$0.003 & 0.199$\pm$0.004 \\
  $\mathbb{P}_{\rm theor}$ & 0.2435 & 0.222 & 0.209 & 0.20 \\
  \hline
\end{tabular}
\caption{Experimental and theoretical values of the linear
polarization. The theoretical values are computed using optical
data for $T=2400$ K.}\label{comp}
\end{table}

\section{Conclusions}

The fact, contrary to one's intuition of thermal phenomena, that
thermal radiation from incandescent bodies may reveal unexpected
coherent features is nowadays well appreciated. A remarkable
example of this sort is the polarization of the radiation emitted
by a thin incandescent wire, in the direction orthogonal to the
wire, that was reported for the first time by  $\ddot{\rm O}$hman
long ago \cite{ohman}. These initial findings were later confirmed
in a preliminary series of measurements by Agdur et
al.\cite{agdur} with platinum filaments having thicknesses between
one and ten microns. These authors observed an increasing   linear
polarization with decreasing thicknesses and a maximum
polarization of about fifty percent. However, the insufficient
quality of the measurements did not allow for a rigorous
comparison between theory and experiment.  In this paper we have
reported new measurements of the linear polarization of thermal
radiation emitted by incandescent thin tungsten wires, with
thicknesses ranging from five to hundred microns, and temperatures
in the interval from 2600 to 3200 K. For thicknesses in this range
we observe an increasing   linear polarization with decreasing
thickness, in qualitative agreement with the results of
\cite{agdur}. We have compared our measurements with theoretical
predictions, based on the available optical data for tungsten
\cite{roberts}, referring to filaments with temperatures ranging
from 298 K to 2400 K, and we found  very good agreement between
our measurements and the theoretical prediction derived from the
2400 K data. Interestingly enough, for small wire thicknesses,
theory predicts an inversion of the qualitative dependence of the
linear  polarization with wire thickness. As it can be seen from
Fig. 4, we note that for thicknesses smaller than about four
microns, a decrease of thickness is expected to engender a smaller
polarization, contrary to the behavior predicted (and observed by
us) for thicker wires. Unfortunately, we could not perform any
measurements with wires thinner than five microns, in order to
observe this interesting inversion phenomenon.

\section{Appendix}

The optical properties of tungsten, in the wavelength range from
0.365 to 2.65 microns, were measured long ago by Roberts
\cite{roberts}. He showed that the following formula for the
permittivity $\epsilon$, adapted from Drude's well known
expression, adequately fits the data: \be \epsilon=1+\sum_p
\frac{K_{0p} \lambda^2}{\lambda^2-\lambda_{sp}^2+i \delta_p
\lambda_{sp} \lambda}\,-\,\frac{\lambda^2}{2 \pi c
\epsilon_0}\sum_q \frac{\sigma_q}{\lambda_{r q}-i
\lambda}\,,\label{drude}\ee where $\lambda$ is the wavelength in
vacuum, $c$ is the velocity of light and $\epsilon_0$ is the
permittivity of vacuum (in mks units). The  first sum in the
r.h.s. of Eq. (\ref{drude})  represents a bound-electrons
contribution, while the second sum is a free-electron
contribution. If the above equation is extrapolated to very low
frequencies, one obtains the limiting value $\sigma_0=\sum_q
\sigma_q$ for the dc conductivity. For convenience of the reader,
the numerical values of the parameters for a number of
temperatures, as quoted in \cite{roberts}, are reproduced in Table
\ref{optical}.
 The bound-electron contribution at the higher temperatures is
substantially the same as that at 1600 K, and for this reason the
corresponding parameters $K_{0p}, \lambda_{sp}$ and $\delta_{p}$
are not displayed in the last two columns of Table \ref{optical}.
\begin{table}
\begin{tabular}{|c|c|c|c|c|c|}
  \hline
  Temp. & 298 K & 1100 K & 1600 K & 2000 K& 2400 K \\
  \hline
  $\sigma_1$ & 17.50 & 3.50 & 2.14 & (1.58) & (1.19) \\
  $\sigma_2$ & (0.21) & 0.16 & 0.19 & (0.22) & (0.25) \\
  $\lambda_{r1}$ & 45.5 & 9.3 & 6.0 & (4.63) & (3.66) \\
  $\lambda_{r2}$ & (3.7) & $<0.36$ & $<0.36$ & $(<0.36)$ & $(<0.36)$ \\
  $K_{01}$ & 12.0 & 10.9 & 10.9 &  &  \\
  $K_{02}$ & 14.4 & 13.4 & 13.4 &  &  \\
  $K_{03}$ & 12.9 & 12.0 & 12.0 &  &  \\
  $\lambda_{s1}$ & 1.26 & 1.40 & 1.40 &  &  \\
  $\lambda_{s2}$ & 0.60 & 0.57 & 0.57 &  &  \\
  $\lambda_{s3}$ & 0.30 & 0.25 & 0.25 &  &  \\
  $\delta_1$ & 0.6 & 1.0 & 1.0 &  &  \\
  $\delta_2$ & 0.8 & 1.2 & 1.2 &  &  \\

  $\delta_3$ & 0.6 & 1.0 & 1.0 &  &  \\
  $\sigma_1/\lambda_{r1}$ & 0.385 & 0.376 & 0.357 & (0.341) & (0.325) \\
  $\sigma_0$ & 17.7 & 3.67 & 2.34 & 1.80 & 1.44 \\
  \hline
\end{tabular}
\caption{Optical data for tungsten from \cite{roberts}. $(\;)$
indicates tentative estimates. Conductivities ($\sigma_1$, etc.)
are in units of $10^6$ ohm$^{-1}$m$^{-1}$. The dc conductivity is
$\sigma_0$. Wavelengths ($\lambda_{r1}$, etc.) are in
microns.}\label{optical}
\end{table}

\end{document}